\documentclass[useAMS,usenatbib]{mn2e}
\usepackage[utf8]{inputenc}
\usepackage[T1]{fontenc}
\usepackage{epsfig}
\usepackage{graphicx}
\usepackage{graphics}
\usepackage{color}
\usepackage{amssymb}
\usepackage{amsmath}

\title[Ionizing Emission from SN Ia Progenitors]{He II recombination lines as a test
of the nature of SN Ia progenitors in elliptical galaxies}
\author[T.E. Woods and M. Gilfanov]{T.E. Woods$^{1}$\thanks{E-mail:
tewoods@mpa-garching.mpg.de} and M. Gilfanov$^{1,2}$\\ $^{1}$Max Planck Institute for Astrophysics, Karl-Schwarzschild-Str.
1, Garching b. M{\" u}nchen 85741, Germany\\
$^{2}$Space Research Institute of Russian Academy of Sciences, Profsoyuznaya 84/32,117997 Moscow, Russia\\}

\begin{document}

\pagerange{\pageref{firstpage}--\pageref{lastpage}} \pubyear{2012}

\maketitle

\label{firstpage}

\begin{abstract}
To date, the question of which progenitor channel can reproduce the observed rate of Type Ia supernovae (SNe Ia) remains unresolved, with the single and double degenerate scenarios remaining the leading contenders. The former implies a large population of hot accreting white dwarfs with photospheric temperatures of $T\sim 10^5-10^6$ K during some part of their accretion history. We show that in early-type galaxies, a population of accreting white dwarfs large enough to reproduce the SN Ia rate would contribute significantly to the ionizing UV radiation expected from the stellar population. For mean stellar ages $\lesssim$ 5 Gyr, single degenerate progenitors would dominate the ionizing background produced by stars, increasing the continuum beyond the He II-ionizing limit more than ten-fold. This opens a new avenue for constraining the progenitors of SNe Ia, through consideration of the spatially extended low-ionization emission-line regions now found in many early-type galaxies. Modelling the expected emission, we show that one can constrain the contribution of the single degenerate channel to the SN Ia rate in E/S0 galaxies from upper limits on the luminosity of He II recombination lines in the optical and FUV. We discuss future directions, as well as possible implications for the evolution of SNe Ia in old stellar populations. 
\end{abstract}


\section{Introduction}

Type Ia supernovae (SNe Ia) remain without a definitive model for their origin, despite their great importance in the measurement of cosmic distances \citep{Riess98, Perlmutter99} and their vital role in the chemical evolution of the Universe \citep[see e.g.][]{MG86}. The two leading models are the single degenerate scenario \citep{WI73}, in which a single white dwarf (WD) reaches the Chandrasekhar mass ($\rm{M}_{\rm{Ch}}$) through accretion from a red giant or main sequence companion; and the double degenerate scenario \citep{Webbink84, IT84}, in which a SN Ia results from the inspiral and merger of a binary pair of white dwarfs. Unfortunately, unlike the progenitors of Type II supernovae, SN Ia progenitors are too faint to be detectable \citep[to date, e.g.][]{Li11, NVN12} in archival images retrieved after the explosion. Therefore, indirect arguments must be introduced in order to place limits on the likely progenitors. 

The single degenerate channel implies the existence of a large population of steadily accreting WDs. If steady nuclear burning occurs near the surface, each of these should spend some fraction of its accretion history as a strong supersoft X-ray source \citep[SSSs,][]{vdHeuvel92, Kahabka97}. Recently however, the total soft X-ray emission of elliptical galaxies was found to be far too low \citep{GB10}, and the number of observed SSSs in nearby galaxies too few \citep{DiStefano10}, for the SD scenario to hold in its standard form. Population synthesis calculations have also demonstrated the great difficulty of producing a sufficient number of high-mass, accreting WDs to account for a significant fraction of the SN Ia rate \citep[see e.g.][and for an extreme example \cite{Yungelson10}]{Ruiter09}, at least given our current understanding of binary evolution.

As a possible resolution to the deficit of observed supersoft sources, \citet{HKN10} suggested that a WD's photosphere is inflated in a so-called accretion wind state for much of its accretion phase. This shifts the peak of its emission from soft X-rays to the (extreme) UV, where it can not be directly observed due to absorption. However, for mass transfer rates $\la$ a few $10^{-6}M_{\odot}/\rm{yr} $, relevant to this scenario, the accreting WD should remain an incredibly powerful ionizing source with photospheric temperatures on the order of  ~$\ga10^{5}K$ \citep{HKN99}, possibly resembling a Wolf-Rayet star \citep{USS}. 

At least $\sim 40\%$ of nearby early type galaxies outside the Virgo cluster are known to contain a detectable mass  of neutral hydrogen \citep[$M(H I)\sim 10^8-10^9~M_\odot$,][]{ATLAS3D_12}. The morphology of this gas ranges from irregularly distributed clouds scattered throughout the galaxy to regular disks or rings, the latter constituting the majority of H I detections. The H I disks may be confined within the stellar body of the galaxy, or extend to tens of kpc far beyond the stellar light, and have typical H I column densities on the order of $N_{\rm{H}}\sim 10^{20}$ cm$^{-2}$ \citep{ATLAS3D_12}.

In addition to the neutral matter are {\it extended} (well beyond the nucleus) regions of LINER\footnote{Low-Ionization Nuclear Emission-Line Regions \citep{Osterbrock}.}-like line emission \citep[see e.g.][]{SAURON}, indicating the presence of warm (T $\approx$ $10^{4}$ K) ionized gas. This appears to be primarily powered by a diffuse galactic background, rather than being the summation of many overlapping H II regions. Post asymptotic giant branch stars (pAGBs) may provide the dominant contribution to the needed diffuse ionizing background \citep{SAURON, Eracleous10}. The ionizing spectra of pAGBs also produce reasonable agreement with the observed optical line ratios \citep{Binette94, Stasinska08}, and appear to adequately predict the radial dependence of the inferred ionizing flux \citep{Yan12}. 

A population of accreting white dwarfs implied by the SD scenario, with temperatures $\ga 10^5$ K  would, however, introduce a substantially harder component to the ionizing background, in particular beyond the He II photoionizing limit. The presence of such a component would dramatically change the ionization structure of the H II gas and the line content of its emission. The greater flux of high energy ionizing photons will have a pronounced effect on several characteristic line ratios \citep[][]{Rappaport94}. Most importantly, the total luminosity reprocessed into recombination lines of He II should be greatly enhanced. These lines are characteristic of the youngest starbursts (the result of ionization by WR stars), and are not normally observed in old stellar populations outside the innermost nuclear regions. The strongest He II recombination line which is not capable of ionizing hydrogen, and therefore is not heavily absorbed in the ISM, is the $n=3\rightarrow 2$ transition at $\approx 1640$\AA. In the optical band, the strongest line is the $n=4\rightarrow 3$ transition at $\approx 4686$\AA.  The strength of these lines (or their absence) in the optical and UV spectra of  early type galaxies can be used to limit the total population of such purported SN Ia progenitors, direct emission from which is unobservable, due to interstellar absorption. Furthermore, contrary to upper limits from X-ray observations, such a method should in principle be strengthened if emission is primarily absorbed near the source, given the higher covering fraction and ionization parameter.

With this in mind, we explore the consequences of introducing a population of SD SN Ia progenitors, which spend some fraction of their accretion history at varying temperatures typical of nuclear-burning WDs. 
We proceed as follows: in $\S2$, we estimate the ionizing background expected from the passively evolving (SFR $\approx$ 0) stellar population in early-type galaxies, and its modification due to the contribution of a putative population of accreting white dwarfs envisaged by the  SD SN Ia scenario. In $\S 3$ we consider the ionization balance in the photo-ionized gas and model the emitted spectra of these extended nebulae.  In particular, we investigate how the total expected luminosity of H and He II recombination lines should vary with age of the underlying stellar population. We demonstrate how one can then constrain the maximum luminosity which may be radiated by a population of SD progenitors at a given temperature using the predicted and observed He II 4686\AA\ flux (or for the latter, upper limits). Finally, we discuss observational prospects and what implications our results have for the possible role of the SD SN Ia channel, as well as whether there may be any caveats to the limits given here.

\section{The Ionizing Background in Early-Type Galaxies}

\subsection{Post Asymptotic Giant Branch Stars}

In order to model the ionizing background from the underlying stellar population, we make use of the spectral evolution calculations of \cite{BC03}. For simplicity we model all passively evolving galaxies as having stellar populations born in a single burst of star formation, with initial masses distributed following \cite{Chabrier}. Though certainly an oversimplification, focussing on such simple populations allows us to emphasize the importance of any age-dependence for the ionizing background. In practice, it is then straightforward to contruct more complex star-formation histories. 

In this work, we take Z = 0.05 ($\approx$ 2.5$\rm{Z}_{\odot}$) as the typical metallicity of stellar populations in early-type galaxies, consistent with the value assumed in the "baseline" measurement of the SN Ia rate in \cite{Totani08}. Such a metallicity is well in line with observations of old stellar populations. Although this gives a higher K-band luminosity per unit mass at any given age (increasing the predicted number of SN Ia progenitors), the ionizing continuum from a simple stellar population (SSP) is also generally higher for higher metallicities, minimizing the importance of the stellar population's metallicity (with the continuum varying by only $\approx$ 10 -- 20\%, except at the earliest ages). At 1Gyr, the ionizing continuum is roughly the same for 2.5$Z_{\odot}$ as for $Z_{\odot}$, but the SN Ia rate would be $\sim 40\%$ higher for 2.5$Z_{\odot}$.  

\subsection{Single Degenerate Progenitors}\label{SD}

We can also make an estimate of the expected emission from SD SN Ia progenitors. In the standard SD scenario, the total mass of the WD must reach $M_{\rm{ch}}$ $\approx$ $1.4M_{\odot}$ through accretion from either a main sequence or red giant companion. In order to avoid an accretion-induced collapse, the accretor must be a carbon-oxygen (CO) WD. The maximum mass with which a CO WD can be born is presently estimated to be approximately $\sim 1.1 M_{\odot}$ \citep{Umeda99}, therefore a SN Ia progenitor must accrete {\it at least} $\Delta M\sim 0.3 M_{\odot}$  before reaching $M_{\rm{Ch}}$. Note however that the distribution of WD initial masses falls steeply with greater mass, from a peak around $\rm{M}_{\rm{WD}}$ $\approx$ 0.6$\rm{M}_{\odot}$,  therefore the mean accreted mass per one type Ia supernova can be expected to typically greatly exceed the value quoted above.

The luminosity of an accreting WD is dominated by the energy release from nuclear burning of accreted matter so long as the mass transfer rate exceeds the minimum limit:

\begin{equation}
 \dot M_{\rm{steady,min}} \ga 3.1 \cdot 10^{-7} (M_{\rm{WD}}/M_{\odot} - 0.54)M_{\odot}/\rm{yr}
\end{equation}
\citep{Nomoto07}. Below this, nuclear burning occurs only in unsteady bursts, giving rise to novae. The retention fraction of matter in this case remains highly uncertain; even optimistic binary evolution models suggest $\Delta m_{\rm{WD}}$ $\lesssim$ $0.1 M_{\odot}$ may be accreted in this low-$\dot M$ regime \citep[see e.g.][]{HKN10}. In such a case, SN Ia progenitors would necessarily undergo many recurrent novae, in conflict with the observed rate of novae in nearby galaxies \citep{Gilfanov11}.

Assuming the steady burning condition is met, the luminosity of the accreting WD is then:

\begin{equation}
L_{nuc} = \epsilon _{\rm{H}}\chi \dot M \label{LWD}
\end{equation}

\noindent where $\epsilon _{\rm{H}}$ is the energy release per unit mass of hydrogen ($\sim 6\cdot10^{18}$ erg/g), and $\chi$ is the mass fraction of hydrogen ($\sim 0.72$). Helium burning is significantly less efficient and contributes to the luminosity only on the order of $\sim 10\%$. 

For any given WD mass, there is also a maximum accretion rate for steady burning; when the luminosity appproaches that of a red giant the burning luminosity in eq. \ref{LWD} is limited by \citep{Nomoto07} 

\begin{equation}
\dot M_{\rm{RG}} \approx 6.7 \cdot 10^{-7} (M_{\rm{WD}}/M_{\odot} - 0.45)M_{\odot}/\rm{yr} 
\end{equation}
Above this limit, it is not immediately apparent what becomes of the excess matter. It is possible that a thick envelope will form around the accreting WD, effectively returning it to a red giant state. However, if this is the case, it is also unlikely that the WD will be able to accrete and retain sufficient matter to produce a SN Ia \citep[e.g. ][]{Cassisi98}. Another possibility is that for any mass transfer rate greater than this, the excess matter is lost in a fast wind driven from within the WD's now significantly expanded photosphere \citep{HKN99}. The possible accretion states are then:

\begin{itemize}
\item $\dot M$ $>$ $\dot M_{\rm{RG}}$ \hfill Envelope or wind

\item $\dot M_{\rm{steady,min}}$ $<$ $\dot M$ $<$ $\dot M_{\rm{RG}}$ \hfill Steady burning

\item $\dot M$ $<$ $\dot M_{\rm{steady,min}}$ \hfill Unstable, nova flashes
\end{itemize}

In the SD scenario, the total number of SN Ia progenitors within a galaxy with instantaneous SN Ia rate $\dot N_{\rm{Ia}}$ is given by \citep{Gilfanov11}

\begin{equation}
N_{\rm{progenitors}} \approx \dot N_{\rm{Ia}} \cdot \Delta M /\dot M
\end{equation}

We compute the specific SN Ia rate per unit K-band luminosity from the delay-time distribution found by \cite{Totani08} in galaxies with old stellar populations

\begin{equation}
\dot N_{\rm{Ia}} = 0.57 (t/\rm{Gyr})^{-1.11} Sne/century/10^{10} L_{\rm{K},\odot}
\end{equation}
Here we use their fit obtained for Z = 2.5$\rm{Z}_{\odot}$ and a Chabrier IMF. 
Note that in the study of \cite{Totani08}, they take the rate per unit K-band luminosity found by \cite{Mannucci05} for E/S0 galaxies as the value at $t$ = 11 Gyr. 

If we assume that throughout the growth of the accreting WD, the mass transfer rate lies somewhere in the range $10^{-7}$ $M_{\odot}/\rm{yr}$ $\lesssim$ $\dot M$ $\lesssim$ $10^{-6}$ $M_{\odot}/\rm{yr}$, then the number of (either steadily or unsteadily) nuclear burning WDs in a 10 Gyr-old elliptical with a K-band luminosity $10^{11}$ $L_{\rm{K,}\odot}$ is accordingly $10^{3}$ $\lesssim$ $N_{\rm{progenitors}}$ $\lesssim$ $10^{4}$.

While their total number may be difficult to pinpoint due to the uncertain (yet certainly varying) accretion rate, there is far less freedom in the total bolometric luminosity emitted by a population of SD progenitors. We can estimate this luminosity as 

\begin{equation}
L_{\rm{tot,SNIa}} \approx N_{progenitors} \cdot L_{nuc}  = \epsilon _{\rm{H}}\chi \Delta M_{\rm{Ia}} \dot N_{\rm{SN Ia}}
\label{eq:ltot}
\end{equation}

\noindent Note that the mass transfer rate cancels out in this estimation; whether a few nuclear-burning WDs accrete very quickly until reaching $M_{\rm{Ch}}$, or many WDs accrete very slowly, a certain amount of mass must be processed in the SD scenario at any instant, fixed by the SN Ia rate. 
If, for a given SD progenitor scenario, the accreting component is X-ray or UV-luminous only for some subset of its entire accretion history, we can compute the total luminosity from eq.(\ref{eq:ltot}), replacing the total $\Delta M$ with the amount of mass $\Delta m_{\rm{i}}$ accreted during that phase.

Choosing a particular model for the spectra of SN Ia progenitors, we can then estimate the energy output per unit wavelength of their entire population in any galaxy, from the latter's age and K-band luminosity:

\begin{equation}
L_{\lambda\rm{, net}} = \dot N_{\rm{Ia}}\int _{\rm{M}_{\rm{i}}}^{\rm{M}_{\rm{Ch}}} \frac{L_{\lambda}(M, \dot M)}{\dot M}\rm{d}M\label{eq_L_SD}
\end{equation}
The mass transfer rate $\dot{M}$ varies over the course of the WD's accretion phase, and the emission spectrum $L_{\lambda}(M, \dot M)$ generally depends on the white dwarf mass and the accretion rate. 
 
The case of supersoft sources (steady nuclear burning WDs)  was studied extensively by \citet{Rauch10}. It has been shown that, despite a number of strong atmospheric absorption lines from various elements,  the overall shape of their ionizing continuum does not deviate too strongly from that of a blackbody whose temperature is close to the effective SSS temperature. The latter is given by

\begin{equation}
T_{\rm eff}=\left(\frac{L_{nuc}}{4\pi R_{WD}^2 \sigma_{SB}} \right)^{\frac{1}{4}}
\approx 5.3\cdot 10^5\   \dot{M}_{-7}^{1/4} \ R_{-2}^{-1/2} ~K
\end{equation}

\noindent where $\dot{M}_{-7}$ is the mass accretion rate in units of $10^{-7}~M_\odot/$yr and  $R_{-2}$ is the white dwarf radius in units of $10^{-2}~R_\odot$. Thus, typical SSS temperatures lie in the range $\sim 2$ -- $7$ $\cdot$ $10^{5}K$ \citep{vdHeuvel92}. 

For the optically-thick wind case however ($\dot M$ $>$ $\dot M_{\rm{RG}}$), the expected observable appearance is less certain.  For expected mass transfer rates ($\sim 10^{-6}\rm{M}_{\odot}/\rm{yr}$) the effective temperature should be $\sim$ 2 $\cdot$ $10^{5}K$ for WDs of mass $\ga 0.9\rm{M}_{\odot}$ \citep{HKN99}.  One can demonstrate that for these parameters, the nuclear-burning WD should be able to fully ionize hydrogen and helium beyond the wind's photosphere (see appendix \ref{no_edge}); therefore in the energy range of interest the spectrum should be at least roughly  represented by a black body. In particular, no photoabsorption cutoff at the  He II 54.4eV edge should be expected \citep[cf][]{Rauch03}.
Therefore in the following calculations we assume blackbodies of fixed temperatures to represent the emission spectra of SD SN Ia progenitors. One can then test any SN Ia progenitor channel through superposition of the appropriate results, for differing $L_{\rm{Ia}}$ and $\rm{T}_{\rm{eff}}$.

\subsection{The Ionizing Radiation Field in Ellipticals}\label{flux}

\begin{figure}
\begin{center}
\includegraphics[height=0.25\textheight]{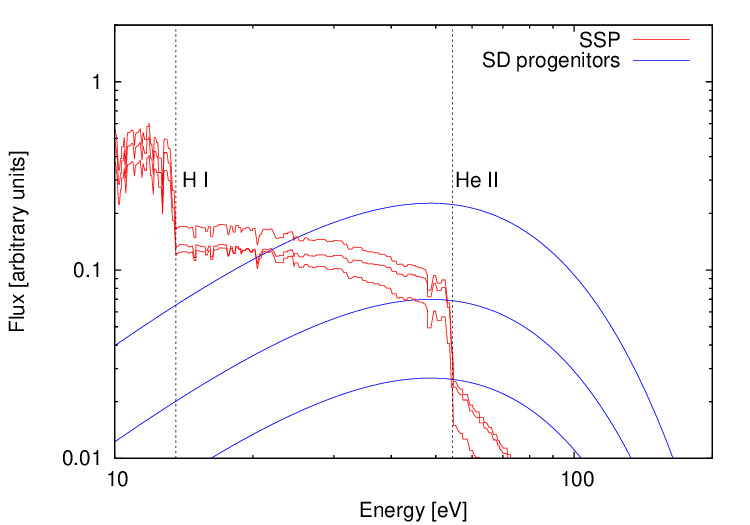}
\caption{Comparison of the emitted spectrum at 3, 6, and 10 Gyr (upper to lower lines) from the normally-evolving stellar population (red) and from a population of SD SN Ia progenitors (blue) assuming that $\Delta M=0.3 M_\odot$ of material is accreted with an effective temperature of $T=2\cdot 10^5$ K, as described in $\S$2.3. 
 \label{comparison}}
\end{center}
\end{figure}

We can now compare the anticipated ionizing emission from normally-evolving SSPs with that from a putative SD progenitor population. Shown in Fig. \ref{comparison} is the emitted spectrum predicted for 3, 6, and 10 Gyr-old populations, overlayed with that from a hypothetical SD channel which processes $\Delta M$ = $0.3M_{\odot}$ at an effective temperature of $\approx$ 2 $\cdot$ $10^{5}K$ (near the lower bound for observed temperatures in symbiotic SSSs, and consistent with what would be expected during the wind phase). Above the He II-photoionizing limit, we see that the contribution from SD progenitors dramatically hardens the spectrum, and begins to dominate even the H-ionizing continuum for young, passively-evolving stellar populations.

This is further illustrated in Fig.2 where we plot the integrated total H- and He II-ionizing photon emission for two cases: photoionization by a SSP, and by an example SD progenitor population. The He II-ionizing luminosity is increased by a factor of $\sim 5$ at $\sim$ 10 Gyr, rising to roughly two orders of magnitude for the youngest populations. There is an accompanying, less significant boost in the H-ionizing luminosity, though this rises to an order of magnitude at the youngest ages.
Clearly, this will have a strong impact on the observed emission lines in any nebulae ionized by the diffuse background produced by such a population. In particular, this suggests that a population of accreting WDs sufficient to match the SN Ia rate should make their presence clear through the observation of He II recombination lines in the spectra of low-ionization emission line regions in passively-evolving galaxies. In order to quantify this, we turn now to modeling these nebulae.

\begin{figure}
\begin{center}
\includegraphics[height=0.25\textheight]{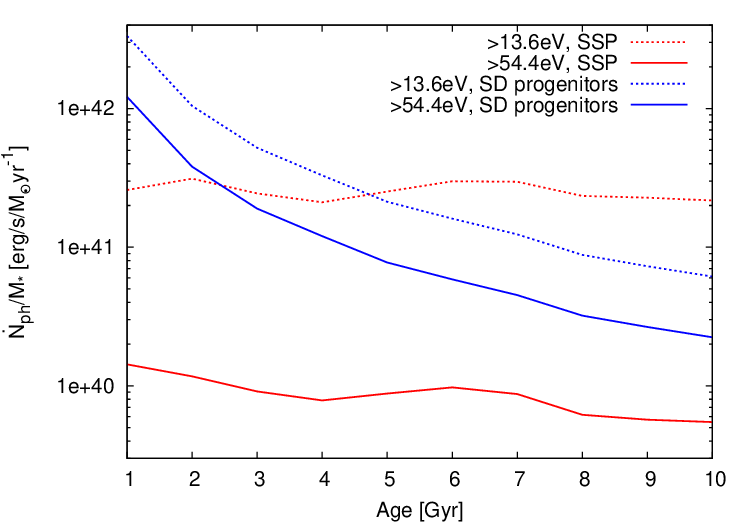}
\caption{Comparison of the H I (dotted lines) and He II (solid lines) ionizing photon luminosity per unit mass from pAGB stars (red lines) with that from a population of SD progenitors (blue lines) as a function of stellar age (from initial starburst). The latter was computed assuming that $\Delta M=0.3 M_\odot$ of material is accreted with an effective temperature of $T=2\cdot 10^5$ K.}
\label{ionizing_luminosities}
\end{center}
\end{figure}

\begin{figure*}
\begin{center}
\hbox{
\includegraphics[height=0.25\textheight]{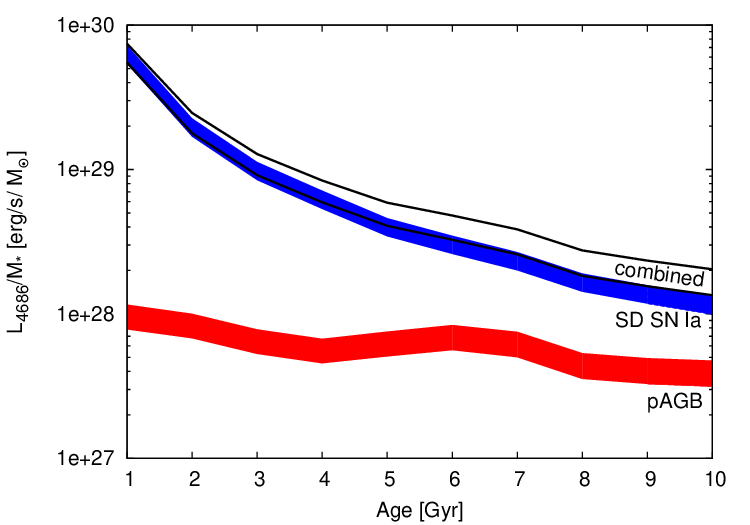}
\includegraphics[height=0.25\textheight]{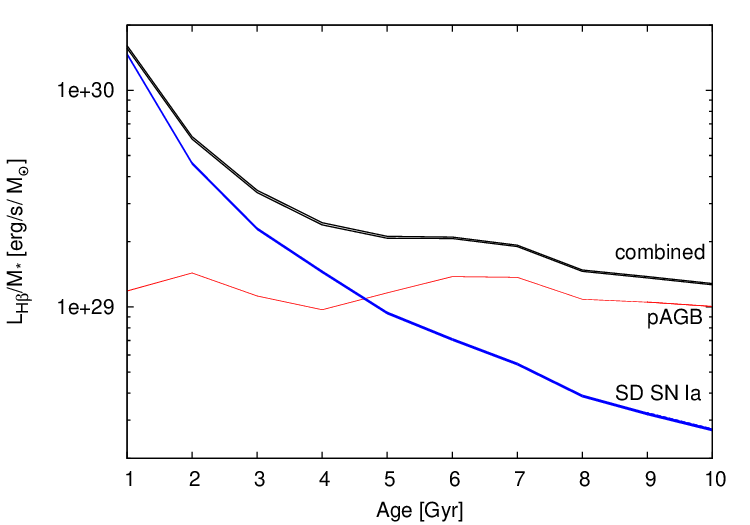}}
\caption{He II 4686\AA\  ({\em left}) and H$\beta$  ({\em right}) luminosity per unit mass assuming ionization by  pAGB stars alone (red), ionization by a population of  SD SN Ia progenitors (blue), and ionization by their combined luminosity (outlined in black), as a function of stellar age (from initial starburst). Here we use the same example parameters for the SD channel as in $\S$ 2.3. In this and all further figures, the thickness of lines shows the effect of the uncertainty in ionization parameter (Section \ref{sec:mappings}). This is negligible for the H$\beta$ line. \label{fig:lines}}
\end{center}
\end{figure*}

\section{Constraining the Characteristics of SN Ia Progenitors}

\subsection{Modelling Low-Ionization Emission-line Regions}
\label{sec:mappings}

Typically, the luminosity in any recombination line can be estimated analytically based on the total ionization rate and branching ratios for recombination \citep[e.g.][]{Osterbrock}. However, the extended emission-line regions of early-type galaxies are observed to be in a very low state of ionization \citep{Binette94}, such that this is not sufficient in computing the luminosity of He II recombination lines (see appendix \ref{HeII}). Therefore, to determine the effect on the nebular emission of introducing a previously unaccounted-for ionizing stellar sub-population, we make use of the photoionization code MAPPINGS III \citep[e.g.][]{Kewley01,Groves04}. 

The code allows us to model the emission lines resulting from an incident ionizing background within a 1-D cloud (where we assume plane-parallel geometry). The incident flux is prescribed by an input spectral shape, normalized by the ionization parameter $U = \dot Q_{\rm{ph}}/(n_{\rm{H}}\rm{c})$ (where $\dot Q_{\rm{ph}}$ is the ionizing photon flux and $n_{\rm{H}}$ is the hydrogen density at the outer face of the slab). Previous studies \citep{Binette94,Stasinska08,Yan12} have found that one requires $-4$ $\lesssim$ $\log (U)$ $\lesssim$ $-3.5$ given photoionization by stellar sources, in order to produce the observed LINER-like emission-line ratios.

We assume a fixed hydrogen density of $n_{\rm{H}}$ = $100\rm{~cm}^{-3}$, consistent with the measured ratio of S II 6717\AA/6731\AA\ in these nebulae \citep[e.g.][]{Yan12}.  Metallicities of the warm ISM in passively-evolving galaxies have been estimated from the Oxygen abundance, with \cite{AB09} finding a mean value of $Z_{\rm{Oxygen}}$ $\approx$ $Z_{\rm{Oxygen},\odot}$ in the 7 galaxies in their sample. Therefore, in this work we assume solar metallicity \citep{AG89} for the warm phase ISM in our calculations. \cite{Annibali10} found a broader range of $0.25Z_{\odot}$ $<$ $Z$ $<$ $2Z_{\odot}$, using a much larger sample of galaxies. However, since we focus on recombination lines of He II, the metallicity of the gas has only a relatively small effect \citep{Binette94}. For similar reasons, we ignore dust in our calculations, although it has been found to accompany the warm and neutral gas in elliptical galaxies \citep{SAURON}. The inclusion of dust in our models however introduces only very minimal reddening \citep{Binette94}, principally effecting forbidden lines through depletion of the gas phase metallicity. 

\begin{figure}
\begin{center}
\includegraphics[height=0.25\textheight]{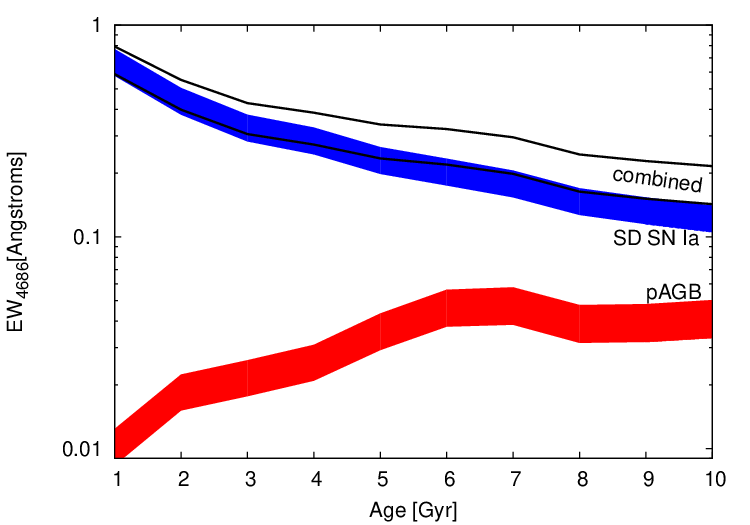}
\caption{Equivalent width of He II 4686\AA\ assuming SSPs as computed by \protect\cite{BC03}. Red denotes the case for ionization by pAGBs only, blue by our standard case SD progenitors, and outlined in black is the EW(4686\AA) assuming ionization by both populations. Here we assume again our standard SD case ($\Delta M=0.3 M_\odot$ of material is accreted with $\rm{T}_{\rm{eff}}$=2$\cdot 10^{5}$ K).}
\label{EW_4686}
\end{center}
\end{figure} 

\begin{center}
\begin{figure}
\includegraphics[height=0.25\textheight]{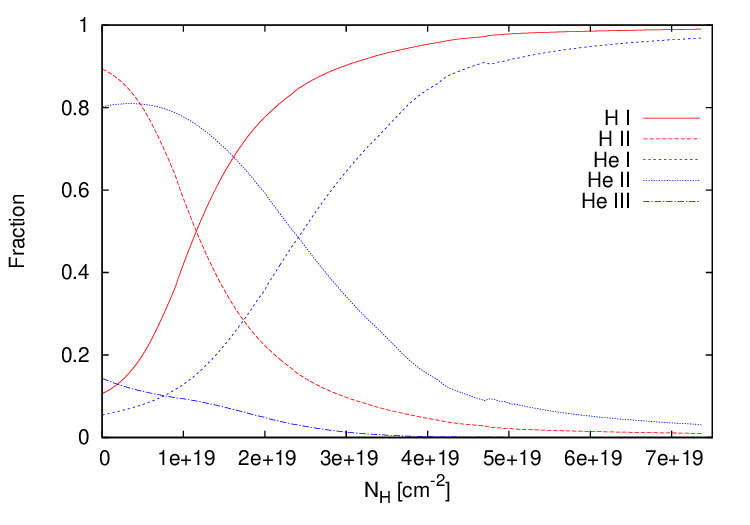}
\caption{Ionization structure for a model nebula assuming plane-parallel geometry, log(U) = --4, and photoionization by a 5 Gyr-old SSP with the contribution of the example SD progenitor population from the text.}\label{slab}
\end{figure}
\end{center}

The gas temperature is computed self-consistently, determined primarily by the spectrum of ionizing radiation for a given density and composition of the nebula.  Along the spatial coordinate, calculations are run until the hydrogen ionization fraction drops below $10^{-2}$. For the parameter range considered here this typically occurs at the depth corresponding to  $N_{\rm{H}}\sim 10^{20}$ cm$^{-2}$. The typical ionization structure for hydrogen and helium obtained in one of the runs is shown in Fig. \ref{slab}. The typical column density of the neutral gas observed in the 21 cm line is of roughly the same order \citep{ATLAS3D_12}. Therefore the standard assumption that nebulae in low-ionization emission-line galaxies are ionization-bounded remains justified, though Str{\"o}mgren boundaries may overlap for relatively high ionization parameter. Note that in Fig. \ref{slab} it is clear all He II recombination emission must principally originate from a relatively thin outer layer. 

Finally, note that unless otherwise stated, we assume a covering fraction ($\rm{f}_{\rm{c}}$) of unity in computing the luminosity of any line. This can be scaled as appropriate given the inferred geometry of the gas; for the case of a smoothly distributed disk (see $\S$ 2.1), we would expect the covering fraction to be $\approx$ 1/2.

\subsection{Recombination Lines of He II}

In the context of this study, the two most important recombination lines of He II are:
\begin{enumerate}
\item the  $n=4 \rightarrow 3$ transition at $\approx$ 4686\AA\ -- the strongest He II line in the optical band 
\item the  $n=3 \rightarrow 2$ transition at $\approx$ 1640\AA\  -- the strongest  He II line longward of 912\AA\ which is not capable of ionizing hydrogen in the ground state, and therefore is not heavily absorbed by the ISM.
\end{enumerate}

In the case of photoionization by the ``normally-evolving'' stellar population, the He II ionization rate and luminosity in any recombination line of He II will remain small (Fig. \ref{comparison}) due to the strong cutoff in pAGB spectra shortward of 228\AA\ \citep{Rauch03}. As the total ionizing luminosity remains relatively constant for SSPs older than $\sim$ 1 Gyr (Fig. \ref{ionizing_luminosities}), the recombination line luminosities do not exhibit any strong trend with age. In particular, the luminosity of any He II recombination line falls slowly by a factor of $\approx$ 2 from 1 -- 10 Gyr (see Fig. \ref{fig:lines} {\it left} for the case of 4686\AA\ emission).

With the introduction of a SD progenitor population (as given in $\S$ 2.3), the He II ionization rate and the luminosity of its recombination lines  is greatly enhanced over the SSP-only case. Due to the $\sim t^{-1}$ dependence of the SN Ia rate, the recombination line luminosity rises steeply with decreasing mean stellar age, from a factor of $\sim$ 3 increase over the SSP-only case at 10 Gyr to $\sim$ 60 for 1 Gyr old populations. We also plot in Fig. \ref{fig:lines} the expected He II 4686\AA\ luminosity given ionization by  SD progenitors alone, showing that the contribution of the SSP does not significantly affect the He II line luminosities. If there are  far fewer pAGB stars then presently predicted by population synthesis \citep{Brown08}, SD progenitors alone would still remain an important ionizing source and our predictions hold nearly unchanged (recall Fig. \ref{ionizing_luminosities} above). 

We can also eliminate any dependence on the total stellar mass by instead considering the He II 4686\AA\ equivalent width (EW(4686\AA)), shown in fig. \ref{EW_4686} for $\rm{f}_{\rm{c}}$ = 1 (again, the EW scales linearly with $\rm{f}_{\rm{c}}$ and can be adjusted accordingly).

Note that there is a range in the predicted He II 4686\AA\ luminosity for all ages. This is a direct result of the range in plausible values of the ionization parameter; for low-ionization nebulae the flux of He II recombination lines has a direct dependence on U (see appendix \ref{HeII}). 

\begin{figure}
\includegraphics[height=0.25\textheight]{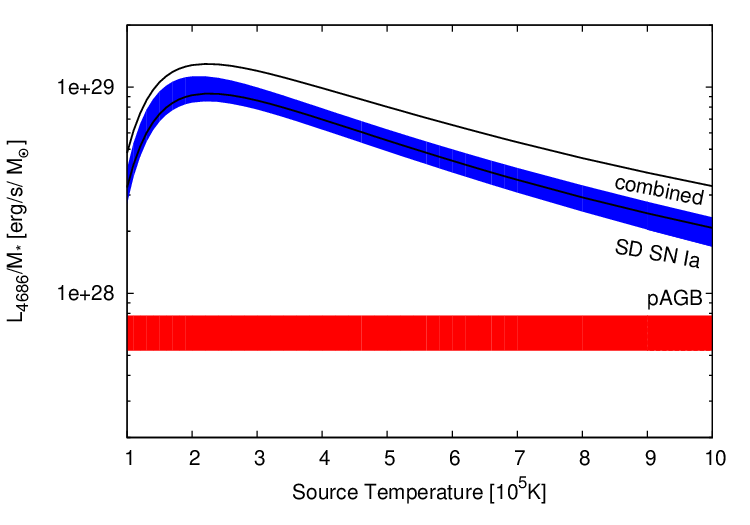}
\caption{Dependence  of He II 4686\AA\  specific luminosity on the  assumed WD photospheric temperature, for $\Delta M$ = 0.3$\rm{M}_{\odot}$ and an age of 3 Gyr. Blue denotes ionization by SD progenitors alone, and outlined in black is the luminosity assuming ionization by SD progenitors combined with a SSP at 3 Gyr. Also shown is the predicted He II 4686\AA\ flux without the influence of any possible SD SN Ia progenitors (red).}\label{lines_T}
\end{figure}

We find similar results for a range in photospheric temperatures spanning that expected for (steadily) nuclear-burning WDs. Plotted in Fig. \ref{lines_T} is the predicted luminosity of the He II 4686\AA\ line per unit mass for varying source temperature, assuming a 3 Gyr-old stellar population. With the inclusion of a SD progenitor population, the expected He II 4686\AA\ line luminosity is enhanced by a factor of 4 (3) for $\rm{T}_{\rm{eff}}$ $\approx$ $10^{5}K$ ($10^{6}K$). This boost is much greater for all intervening temperatures, with a maximum factor of $\approx$ 20 at $\rm{T}_{\rm{eff}}$ $\approx$ 2.5 $\cdot$ $10^{5}K$.

For low-density photoionized gases (with $\rm{T}_{\rm{gas}}$ $\approx$ $10^{4}K$), the luminosity of  the 1640\AA\ line is roughly 6.15 times greater than that in 4686\AA\ emission \citep{Osterbrock}. Given that the FUV continuum in elliptical galxies is nearly 2 orders of magnitude lower than in the optical, the line should also be quite conspicuously prominent. This suggests that FUV spectra available from GALEX may also be quite useful in constraining the hardness of the ionizing spectrum in low-ionization emission-line regions. Unfortunately, there remains a dearth of such galaxies with available FUV spectra \citep{GALEX_Catalog}. We leave further discussion of constraints from the 1640\AA\ line to a future work.

\subsection{H$\beta$ and Diagnostic Line Ratios}
\label{sec:lratio}

As with He II, in the case of pAGB-ionization only, the total luminosity of any recombination line of hydrogen (per unit stellar mass) should remain roughly constant at $\ga1$ Gyr.  This is illustrated in the right panel in Fig. \ref{fig:lines} for the case of H$\beta$. However, in the same figure we see that with the addition of a SD progenitor population, the situation is much different. In this case, we expect an  enhancement by up to an order of magnitude in the flux of H$\beta$ line emission per unit stellar mass for starburst ages of $\la 4$ Gyrs.  

\begin{figure}
\begin{center}
\includegraphics[height=0.25\textheight]{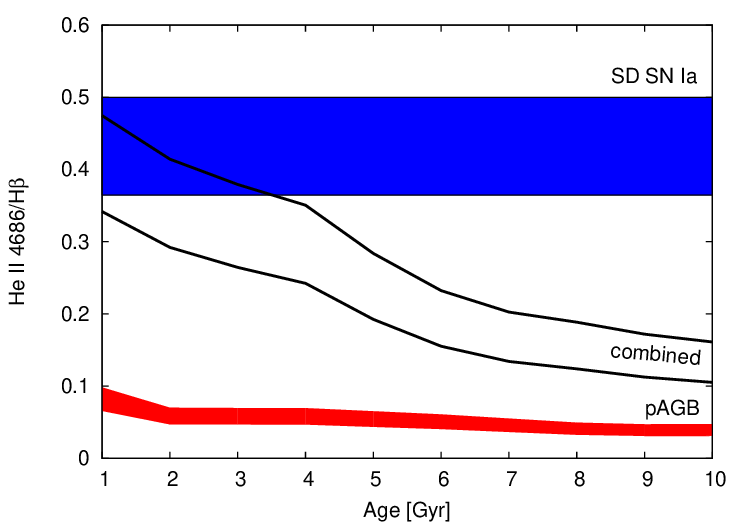}
\caption{The ratio He II 4686\AA/H$\beta$ as a function of stellar age (from initial starburst) for ionization by  pAGBs only (red), a population of  SD progenitors only (blue), and their combined luminosity (outlined in black). Here we use the same example parameters for the SD channel (assuming $\Delta m = 0.3M_{\odot}$, T = 2 $\cdot$ $10^{5}K$). \label{HeIIHb}}
\end{center}
\end{figure}

In principle one can avoid any dependence on the covering fraction if we normalize the He II 4686\AA\ flux to any recombination line of Hydrogen, such as H$\beta$. Though not the strongest optical recombination line of Hydrogen (H$\alpha$/H$\beta$ $\approx$ 3), H$\beta$ is not too faint to be detectable in moderately nearby galaxies. At the same time, it is not too greatly separated from the 4686\AA\ line, minimizing the importance of extinction in computing their ratio. The predicted He II 4686\AA/H$\beta$ ratio as a function of age is given in Fig. \ref{HeIIHb}. As one can see, the He II 4686\AA/H$\beta$ ratio changes from $\sim 0.05$ to $\sim 0.4-0.5$ between the SSP-only and SD-only limits.

However, there are a few difficulties with using the  He II 4686\AA/H$\beta$ ratio to constrain the population of SD progenitors.   Firstly, SD progenitors  will also contribute to the ionization of hydrogen, and thus to H$\beta$ emission (Fig. \ref{fig:lines}). Of course, this is accounted for self-consistently  in calculations presented in Fig. \ref{HeIIHb}, but it requires precise knowledge of the spectral shape of the SD progenitors down to to the hydrogen ionization limit of 13.6 eV. The curves in Fig. \ref{HeIIHb} were computed assuming a black body spectrum, which may be only approximately true for rapidly-accreting WDs (see appendix \ref{no_edge}). Secondly, the line ratio is quite sensitive to ionization parameter for such low values (for the same reasons as discussed in appendix \ref{HeII}). This, however, can be estimated from the nearby [O III] 5007\AA\ line; specifically, its ratio to H$\beta$ \citep[which increases monotonically with U for the range in U considered, see e.g. ][]{Binette94}. Finally, there may be H$\beta$ ``contamination'' from other ionizing sources, such as low-luminosity AGN (LLAGN). Such objects likely provide the dominant ionizing source in the nuclear regions of many low-ionization emission line galaxies \citep{Eracleous10, Annibali10}. Therefore, it is important that one isolate ``retired'' galaxies \citep{Fernandes11}, in which the aging stellar population plays the dominant role in photoionizing emission-line regions. Significant contamination can occur with the inclusion of galaxies hosting LLAGN \citep{Annibali10}, or regions with strong shocks \citep{SAURON}. Such galaxies can however be excluded with strict selection criteria, as outlined by \cite{Fernandes11}.

\section{Observational Prospects}

\subsection{Optical and UV observations} 

Spectroscopic surveys suggest that between $\sim$ 50$\%$ -- 75$\%$ (varying from those in the field to clusters) of E/S0 galaxies contain extended emission-line regions. Despite extensive study, no detection of the He II 4686\AA\ line has yet been reported. Moreover, its presumed intrinsic weakness in passively evolving stellar populations rendered this line uninteresting in spectroscopic studies of these galaxies. It is therefore unsurprising that no attempt to find weak 4686\AA\ line emission in early type galaxies  has been undertaken and, consequently, no upper limits have been reported. However, as we discuss below, detection of the He II recombination line emission in the optical and far-UV bands  should certainly be feasible in the SD scenario. Conversely, upper limits on such emission can be used to constrain the collective nuclear burning rate of the population of accreting white dwarfs in a galaxy, and thus gauge the importance of the SD scenario.

For an early-type galaxy hosting $M(t)$ solar masses in stars of age $t$ Gyrs, our calculations predict that in the SD scenario the luminosity of  the He II 4686\AA\ line is (in the absence of any other ionizing source) roughly:

\begin{equation}
L_{4686}\approx 1.4\cdot 10^{40}\ t_{Gyrs}^{-1.7}\ T_5^{-1}\ f_{\rm{c}}\ \frac{\Delta M}{0.3 M_\odot}
\frac{M(t)}{10^{10}M_\odot} \hskip0.2cm \rm{erg/s}
\end{equation} 

\noindent where $f_{\rm{c}}$ is the covering fraction, and $\Delta M$ is the mass accreted by a typical white dwarf with photospheric temperature $T = T_5\times 10^5$ K. This formula approximates the numerical results presented in Figs. \ref{fig:lines} \& \ref{lines_T} and is valid for $T\ga 2\cdot 10^5$ K. In order to obtain the equivalent width, one can divide the specific luminosity by the specific continuum emission at $\approx$ 4686\AA,  given approximately by $L_{\rm{SSP,}4686} = 9.4\cdot10^{29}t^{-1.04}$erg/s/\AA/$M_{\odot}$. We then find the EW(4686\AA) from SD progenitors should follow: 

\begin{equation}
\rm{EW}(4686\text{\AA}) \approx 1.5\ t_{Gyrs}^{-0.7}\ T_5^{-1}\ f_{\rm{c}}\ \frac{\Delta M}{0.3 M_\odot}\text{\AA}
\end{equation}

For an early type galaxy hosting  a 3 Gyr-old $10^{10}M_{\odot}$ stellar population, with an accompanying SD progenitor population (our standard case from above) and $f_{\rm{c}}$ = 1/2, our calculations predict a He II 4686\AA\ line-luminosity of $\sim 3-5\cdot 10^{38}$ erg/s (for -4 $\leq$ log(U) $\leq$ -3.5).  In the absence of any SD progenitors, the predicted line luminosity will be at most $\sim 4\cdot 10^{37}$ erg/s  (assuming ionization by pAGB stars only, Fig.\ref{fig:lines}), i.e.  $\approx 10$  times smaller. At a distance of 100 Mpc, the expected line flux in the He II 4686\AA\ line is $\approx  4\cdot 10^{-16}$ erg/s/cm$^2$ for the SSP+SD case. 
As a conservative estimate, we can assume here a limit for a confident line detection of $\sim 10^{-16}$ erg/s/$\rm{cm}^{2}$.  Thus, our example galaxy above should be detectable out to $\sim$ 200 Mpc  (or z $\sim 0.05$). Ongoing integral field spectroscopic surveys, such as CALIFA \citep{CALIFA}, should easily achieve comparable or better sensitivity, and therefore will find it well within their grasp to either detect He II 4686\AA\ emission or derive useful upper limits.

Note however that for SDSS galaxies, such an assumption is overly optimistic, as detection is limited by the S/N in the spectra (typically S/N $\approx$ 10 -- 15 per pixel).  On the positive side, however, this is not dominated by systematic errors (e.g. in stellar continuum models; Jonas Johansson, private communication). Therefore, as noted above, stacking spectra may allow for a confident detection of 4686\AA\ emission for much smaller $\Delta M$ or larger distances. Given EW(4686\AA), we can estimate the signal-to-noise ratio needed in order to barely detect the line:

\begin{equation}
\frac{S}{N} = \frac{A/N\sqrt{2\pi}\sigma}{EW(4686\text{\AA})}
\end{equation}

\noindent as given in \cite{SAURON06}. Here we assume an intrinsic broadening $\sigma_{\rm{gas}}$ $\sim$ 150km/s ($\approx 2.3$\AA at 4686\AA) typical of the gas in the central regions of early-type galaxies hosting extended emission-line regions. Assuming a threshold A/N $\approx$ 4 is needed for detection, one would then require S/N $\approx$ 500 in order to detect an EW(4686\AA) $\approx$ 0.05\AA. This would allow for a nominal detection of He II 4686\AA\ from pAGBs for relatively old stellar populations, and enable a strong detection of (or robust upper limit on) any SD progenitor population. In the SDSS sample, such a target S/N should be possible with stacking of $\sim$ 2000 galaxies. 

\begin{center}
\begin{figure}
\includegraphics[height=0.25\textheight]{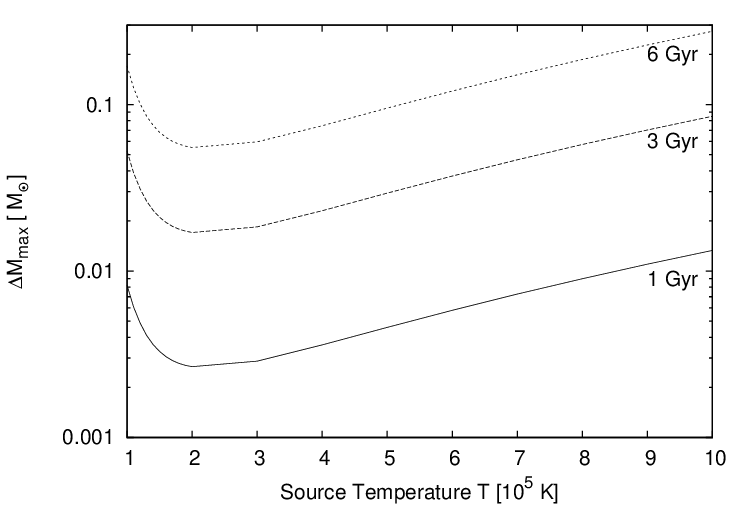}
\caption{Maximum mass which may be accreted by the typical WD in the SD scenario for a given photospheric temperature, given a null detection of the He II 4686\AA\ line with an upper limit of $10^{-16}$ erg/s/$\rm{cm}^{2}$. Here we assume  a 5 $\cdot$ $10^{10}M_{\odot}$ stellar population aged 1, 3, and 6 Gyrs, a distance of 100 Mpc, with a gas covering fraction of 1/2, and log(U) = -4.}\label{sensitivity}
\end{figure}
\end{center}

Fig. \ref{sensitivity} illustrates how an upper limit on the He II 4686\AA\ line can be used to constrain the SD scenario. It shows the upper limit which can be placed on the mass $\Delta M$ accreted by each successful SN Ia progenitor at the given photospheric temperature in the SD scenario, assuming a null detection of the He II 4686\AA\ line with an upper limit of $10^{-16}$ erg/s/$\rm{cm}^{2}$. For illustrative purposes, we assume a 5 $\cdot$ $10^{10}M_{\odot}$ stellar population of varying ages at a distance of 100 Mpc\footnote{Note that this is much too great a mass to be enclosed in an SDSS fiber at this distance, but would be possible with e.g. long-slit or integral field spectroscopy.}, and ionization by a SD SN Ia population only.
The plot demonstrates that even a rather moderate upper limit on the He II 4686\AA\ line emission  
can constrain the mass accreted by each SN Ia progenitor down to $\sim {\rm few} \times 10^{-3}-10^{-1}$ M$_\odot$.
Alternatively, this can also be used to constrain the fraction of SNe Ia produced via the SD channel. Indeed, if we assume some value of $\Delta M$ and $T$, the upper limit on the SD fraction is given by $\Delta M_{\rm max}/\Delta M$.

\begin{figure}
\includegraphics[height=0.25\textheight]{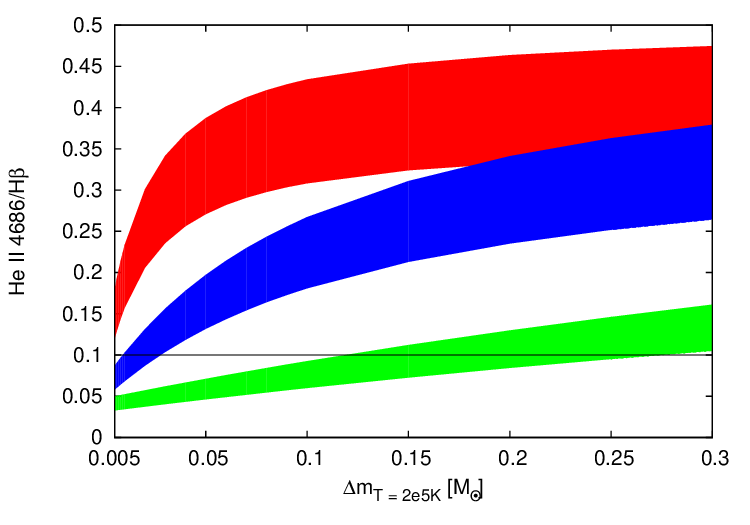}
\caption{ An example of the He II 4686\AA/H$\beta$ line ratio diagnostics. The predicted line ratio for a (1 Gyr -- red,  3 Gyr -- blue,  10 Gyr -- green) SSP+SD population with varying assumed $\Delta$M -- the mass accreted by WDs at the photospheric temperature of  T $\approx$ 2 $\cdot$ $10^{5}K$. Black line shows the line ratio level of 0.1, which should be routinely achievable in spectroscopic observations.}
\label{fig:dm_lratio}
\end{figure}

In their survey of nearby LINERs, the Palomar survey did not detect He II 4686\AA\ line emission, but found upper limits of He II 4686\AA/H$\beta$ $\lesssim$ 10\% \citep{Ho08}. Unfortunately, this number can not be directly used for our purpose, as it applies only to the nuclear regions of mostly old galaxies, where we also expect a central source to play a strong role in the available ionization budget (section \ref{sec:lratio}). However, it does characterize the sensitivity to the He II 4686\AA/H$\beta$ line ratio which can be routinely achieved. The upper limit of $\sim 0.1$ on the He II 4686\AA/H$\beta$ line ratio is large enough not to be in conflict with the pAGB-photoionization model for low-ionization emission-line regions. However, in the absence of contamination by ionizing sources other than pAGB stars and SD progenitors, we should expect to see $0.1$ $\lesssim$ He II 4686\AA/H$\beta$ $\lesssim$ 0.5. Therefore, an upper limit on this ratio of $\sim 0.1$ in extended low-ionization emission line regions will also place meaningful constraints on the SD scenario, as illustrated by Fig. \ref{fig:dm_lratio}.

In the FUV, detection of (or robust upper limits on) the He II 1640\AA\ line is even more promising. The nearby early-type galaxy NGC 3607 ($d\approx 21$ Mpc) has a mean stellar age of $\approx$ 3 -- 4 Gyr \citep{Annibali07}, and is known to host an extended low-ionization emission-line region. We can compute the expected 1640\AA\ line flux from the K-band luminosity \citep[$\approx$ 3 $\cdot$ $10^{10}\rm{L}_{\rm{K,}\odot}$ within 14''x14'' from the 2MASS survey,][]{2MASS}, and the appropriate M/$L_{\rm{K}}$ from \cite{BC03}. Assuming $f_{\rm{c}}$ = 1/2, and our prototypical case for the SD channel, the total luminosity reprocessed into He II 1640 \AA\ emission is $\approx$ 3 $\cdot$ $10^{39}$ erg/s. Using the GALEX Exposure Time Calculator (ETC)\footnote{http://sherpa.caltech.edu/gips/tools/expcalc.html}, we find that this should be detectable with S/N $\approx$ 5 given a $\approx$ 3 ksec observation\footnote{Note that here we report 3 times the value given by the ETC, as recommended in the GALEX Spectroscopy Primer. www.galex.caltech.edu/researcher/gr1\_docs/grism/primer.html}.

\subsection{Possible outcomes} 
 
With the observations described above, there are 3 possible outcomes, all of great interest:

$\bullet$ He II 4686\AA/H$\beta\la$ few $\%$ ($L_{4686}/M_*\la {\rm few}\times 10^{27}$ erg/s/M$_\odot$) at high confidence:  Along with placing tight constrains on the SD scenario, this may pose a  problem for our understanding of ionization by pAGB stars, in particular, their spectra beyond the He II-ionization limit.

$\bullet$ He II 4686\AA/H$\beta$ is detected at $\approx$ few $\%$ ($L_{4686}/M_*\sim 10^{28}$ erg/s/M$_\odot$) for all $t_{\rm{gal}} \gtrsim 1$ Gyr: This would provide strong evidence against a high-temperature SN Ia progenitor population. At the same time it would strongly suggest that pAGB-ionization is the correct model for retired galaxies, especially if there is no strong trend in $L_{H\beta}$/$M_{*}$ with age (cf. Fig. \ref{fig:lines}). 

$\bullet$ The He II 4686\AA\ line is detected, with He II 4686\AA/H$\beta$ $>$ $10\%$ ($L_{4686}/M_*\ga {\rm few}\times 10^{28}$ erg/s/M$_\odot$). This would suggest an alternative to pAGB-only ionization. A significant boost found in He II 4686\AA\ emission for younger stellar populations would conform comfortably with the SD SN Ia scenario.

\subsection{Post-starburst galaxies}

Also of interest are the so-called post-starburst galaxies: early-type galaxies with regions of relatively young mean stellar age and extended nebular emission, wherein a burst of star formation has only recently ceased ($\Delta$t $\lesssim$ 1Gyr). \cite{SAURON} find such galaxies account for roughly 10\% of their sample of nearby E/S0 galaxies. Especially puzzling, were the SD scenario to hold, is the existence of the rare class of E+A galaxies \citep{DG83,YG06, Kaviraj07}. These show evidence of a very young population, yet are without strong emission lines. \cite{SAURON} suggest that these galaxies are within a (possibly very) brief epoch shortly after a sudden burst of star formation has ceased. In this phase, OB stars have left the main sequence ($t > 10^{8}$ years), however there are insufficient pAGB stars to strongly ionize the ISM ($t <$ a few $10^{8}$years). Yet SNe Ia at the shortest delay times can already be expected to detonate in this age range, suggesting that given the SD channel, a large population of accreting WDs would need to be present at this time. These galaxies are understood not to be devoid of gas \citep{Buyle06}, therefore the lack of emission-lines characteristic of ionization by hot accreting WDs is certainly conspicuous.

\section{Conclusions}

We have demonstrated that, for temperatures characteristic of WDs accreting above the steady nuclear burning limit, the SD SN Ia channel implies nuclear-burning WDs should provide a substantial contribution to the He II-ionizing continuum in non-star-forming stellar populations. Their contribution rises dramatically at earlier delay-times, with the enhancement of the He II-ionizing continuum growing to nearly 2 orders of magnitude over that of a SSP alone at $\sim$ 1 Gyr. This provides a unique opportunity to test for the presence of any such population through the detection (or lack thereof) of He II recombination lines. In particular, the luminosities predicted in He II 1640\AA\ and He II 4686\AA\ suggest interesting upper limits should already be possible with GALEX and in the SDSS. Ongoing IFU spectroscopic surveys, such as CALIFA, should also be able to improve upon this further, being capable of isolating emission-line regions with no evidence for any contribution from shocks or a central source.

In the absence of any ionizing sources other than the stellar population, an  upper limit on the He II 4686\AA\ line flux of $\lesssim$ 10\% that of H$\beta$ (or $L_{4686}/M_*\la {\rm few}\times 10^{27}$ erg/s/M$_\odot$) in the low-ionization emission-line regions of many ellipticals should tightly constrain the possible luminosity of any population of hot nuclear-burning WDs. This may also provide a crucial test of models of ionization by pAGB stars in galaxies hosting extended low-ionization emission-line regions, wherein one may expect He II 4686\AA/H$\beta$ $\approx$ a few \%. A non-detection of the He II 4686\AA\ line in a galaxy located within 100 Mpc from the Sun with a rather moderate upper limit of $10^{-16}$ erg/s/cm$^2$  will place an upper limit of $\sim {\rm few}\times 10^{-3}-10^{-1}$ M$_\odot$ on the total mass accreted by each successful SN Ia progenitor at photospheric temperatures in the range $\sim 10^5-10^6$ K.

In the preceeding analysis, we have primarily considered the case wherein SD progenitors account for all SNe Ia. However, one may see from eq. \ref{eq:ltot} that a limit on the total luminosity of any SD SN Ia progenitor population may also be interpreted as a limit on the fraction of the total SN Ia rate which SD progenitors account for. This is in keeping with recent suggestions \citep[e.g.][]{MM12} that there may be evidence for multiple progenitor channels.

\section*{Acknowledgements}

The authors would like to thank the referee for helpful comments and discussion, as well as Jonas Johansson, Marc Sarzi, Pierre Maggi, and Dan Maoz. We thank Brent Groves for making his current version of MAPPINGS III available at http://www.mpia-hd.mpg.de/$\sim$brent/mapiii.html.

\appendix

\section{Do We Expect H/He II Absorption Edges in the Emergent Spectra of Rapidly Accreting WDs?}
\label{no_edge}
An accurate treatment of the radiative transfer physics is necessary in order to build a template of the spectral output of rapidly-accreting WDs, should they undergo an accretion-wind phase \citep{HKN99}. However, for the purpose of this paper, we are principally concerned with whether or not there is a strong He II ionization edge in the emergent spectrum. As a simple check,  we can estimate the number of photons needed to wholly ionize H and He above the photospheric radius of the wind and compare this with  the expected ionizing photon luminosity produced by the nuclear-burning WD.

The total rate of hydrogen recombinations above the wind photosphere is:
\begin{eqnarray}
Q_{\rm{R}} = \int n_{e}n_{H}\alpha _{\rm{A}}(T,H^0)\, dV= \\ \nonumber
\frac{x_e}{4\pi}\left(\frac{\dot M}{\xi\rm{m}_{\rm{H}}\rm{v}}\right)^{2}\int_{R_{\rm{phot}}}^{\infty}\frac{\alpha_{\rm{A}} (T,H^0)}{r^{2}}\, dV
\end{eqnarray}
assuming spherical symmetry. Here $\xi$ = $M_{\rm{wind}}/M_{\rm{H}}$ $\approx 1.44$ for solar abundance, $x_e=n_e/n_H\approx 1.2$ for a fully ionized gas of solar composition,  $m_{\rm{H}}$ is the mass of hydrogen, $\alpha _{\rm{A}}(T,H^0)$ is the Case A recombination rate for hydrogen, $R_{\rm{phot}}$ is the radius of the wind's photosphere ($\sim$ 0.1 -- 1 $R_{\odot}$ in the accretion wind phase), $\dot M$ is the wind mass loss rate, and v is the wind speed (v $\approx$ 1000km/s). Such an approach is similar to that taken by \cite{Cumming96}, with the wind originating at the WD surface rather than a red giant companion. As the wind density is always $\lesssim$ a few $10^{14}\rm{cm}^{-3}$, the coronal/nebular approximation is justified.

As a photoionized gas,  the wind temperature at infinity can be expected to approach $\sim$ $10^{4}K$  so long as there is a flux of ionizing photons. For an {\it extremely conservative} estimate of the photon flux needed to ionize the wind, we may assume simply that $T_{\rm{gas}}$ = $10^{4}K$ for $R_{\rm{phot}}$ $\lesssim$ r $<$ $\infty$. In a more careful treatment, the gas temperature falls quickly from that at the photosphere before levelling off at $T_{\rm{gas}}$ $\approx$ $10^{4}K$. This only lowers the recombination rate $\alpha _{\rm{A}}$, thereby raising the ionization fraction, where the density is highest.

\begin{center}
\begin{figure}
\includegraphics[height=0.25\textheight]{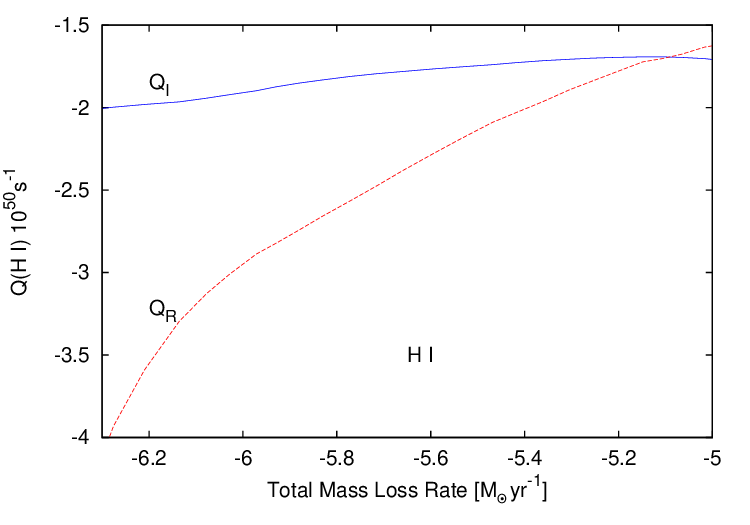}
\includegraphics[height=0.25\textheight]{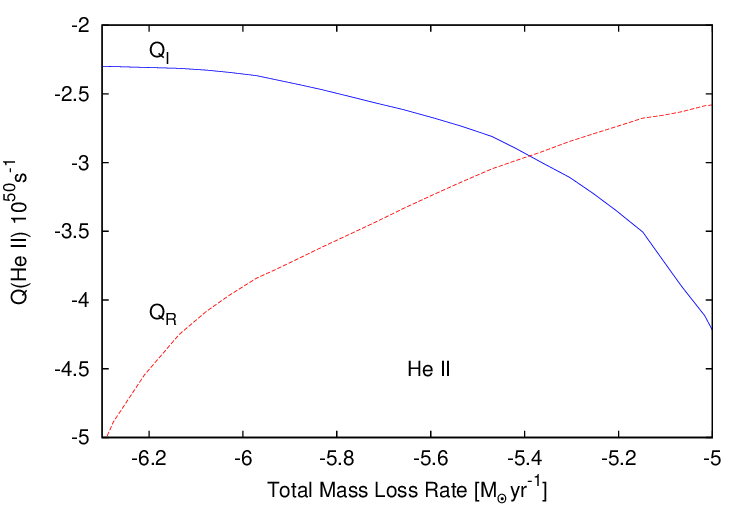}
\caption{Ionizing photon luminosity (solid line, blue in the color version of the plot) and an upper limit on the recombination rate (red dashed line) for hydrogen ({\em Top}) and He II ({\em Bottom}) in a wind-supported extended photosphere of a WD.}
\label{Hrecombination}
\end{figure}
\end{center}

Taking the model of \cite{HKN99}, any quantity of interest in the wind can then be determined from the WD's mass and the mass loss rate of the envelope ($\dot M$ = $\dot M_{\rm{wind}}$ + $\dot M_{\rm{nuc}}$). From their Fig. 3, we take their result for the photospheric temperature as a function of $\dot M$ for a 1 $M_{\odot}$ WD, and set $L_{\rm{nuc}}$ = $\epsilon _{\rm{H}}\chi \dot M_{\rm{RG}}$. From the latter two quantities we can compute the ionizing photon flux at the photospheric radius. In fig. \ref{Hrecombination}, we plot the H-ionizing photon luminosity and the total recombination rate in the envelope versus $\dot M$. We find that the accreting WD's photosphere easily maintains H-ionization in the wind for $\log(\dot{M})\la -5.3$ [M$_{\odot}$/yr].

Given full ionization of hydrogen, a similar equation applies for He II.\footnote{{\it Note that we ignore He I in this check. However, this should not significantly effect the ionization balance, as the He number density is roughly 1/10 that of H, and in the high-density limit more than $\sim$ 1/2 of all He I recombinations produce H-ionizing photons \citep[see][]{Osterbrock}.}} In the lower panel of fig. \ref{Hrecombination}, we compare the expected recombination rate of He II in the accretion-wind atmosphere with the expected He II--ionizing photon luminosity from the wind's photosphere.  Similar to hydrogen, we find that for $\log(\dot{M})\la -5.5$, the ionizing photon rate exceeds that necessary to maintain full He II ionization. Therefore, no significant break should be present at 54.4 eV in the emergent spectra of such WDs. 

Such a mass transfer rate is well within the range relevant to the case of SN Ia progenitors in ellipticals. Indeed, as the mass transfer rate approaches $10^{-5}M_{\odot}$/yr, accretion becomes increasingly inefficient, and required donor masses exceed that available in old populations \citep{GB10}. For example, $\dot{M}_{nuc}/\dot{M}\la 0.1$ for $\dot{M}\ga 10^{-5}$ M$_\odot$/yr \citep{HKN99}, therefore a donor star more massive that $\sim 3-5$ M$_\odot$ is needed in order to provide $\Delta M_{nuc}\sim 0.3-0.5$ M$_\odot$. Such donors are unavailable in the stellar population older that 1(3) Gyrs, where the main sequence turn-off mass is $\sim 2.5$ ($\sim 1.6$) M$_\odot$.

In practice, the wind is not expected to be spherically symmetric, increasing the likelyhood that much of the radiation escapes. Note also that, were the ionizing radiation absorbed near the source, one should expect strong emission lines similar to WR stars \citep{USS}.

\section{He II Recombination Lines in Low Ionization Nebulae}
\label{HeII}

Under the circumstances typically encountered in photoionized nebulae, the flux in any recombination line of $X^{\rm{i}}$ can be computed easily from the incident flux of $X^{\rm{i}}$-ionizing photons, and the ratio of the relevant transition's recombination rate to the total rate of recombinations \citep{Osterbrock}. For the He II 4686\AA\ line, this gives us:

\begin{equation}
L_{4686, \rm{max}} = h\nu _{4686}\frac{\alpha^{\rm{eff}}_{\rm{He II }4\rightarrow3}(\rm{T_{\rm{gas}}})}{\alpha _{\rm{B}}(\rm{T_{\rm{gas}}})}\int^{\infty}_{\nu _{54.4eV}} \frac{L_{\nu}}{h\nu}\rm{d}\nu \label{CaseB}
\end{equation}

However, this assumes that hydrogen is completely ionized, and therefore no He II--ionizing photons are lost in maintaining the ionization of Hydrogen. As pointed out by \cite{ST86} \citep[see also][]{RSF10, Maggi11}, this is no longer valid for high $T_{\rm{source}}$ and low U, in which case He II-ionizing photons are lost in ionizing H I.

In ionization equilibrium, we require that the number of recombinations at any point be equal to the photoionization rate. Ignoring for the moment elements other than hydrogen, we can express the different densities in terms of the particle density using the ionization fraction $x(r)$ $=$ $n_{\rm{e}}(r)/n(r)$ at a depth $r$ within a nebula. We can then find a quadratic expression for $x(r)$:

\begin{equation}
\frac{x^{2}(r)}{1 - x(r)} = \int _{\nu_{0}}^{\infty}\frac{\rm{a}_{\nu}}{\alpha(\rm{T})}\left(\frac{\dot Q_{ph,\nu}}{4\pi r^{2}n_{\rm{H}}}\right)d\nu \label{fraction}
\end{equation}
 
\noindent where $\rm{a}_{\nu}$ is the ionization cross-section, and $n_{\rm{H}}$ is the density of hydrogen. The final bracketed term may be recognized as proportional to the ionization parameter U (when the integration over $\nu$ is carried out). For an incident 54.4 eV photon, the H cross-section is $\sim$ 1/10 that of $\rm{He}^{+}$, however $n_{\rm{H}}$ $\sim$ 10$n_{\rm{He}}$.  Therefore, for relatively low-ionization nebulae, high-energy photons capable of ionizing He II are lost in maintaining the ionization of H I (and He I), greatly diminishing the expected flux of any He II recombination line.

\begin{figure}
\begin{center}
\includegraphics[height=0.25\textheight]{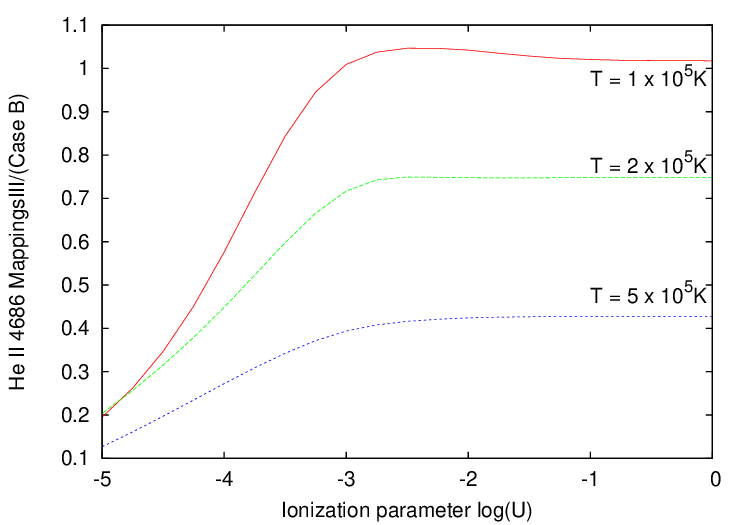}
\caption{The discrepancy between the predicted He II 4686\AA\ flux computed analytically assuming Case B recombination for $T_{\rm{gas}}$ = $10^{4}K$ \citep{Osterbrock}, and the result of an accurate consideration of the ionization structure of the gas using  MAPPINGS III (with gas temperature computed self-consistently, assuming n = 100 $\rm{cm}^{-3}$).  Curves are labeled by temperature of the ionizing source.
\label{He II_dis}}
\end{center}
\end{figure}

From the above, we expect the efficiency in producing He II 4686\AA\ emission to fall off dramatically at low U. It is clear from eq. \ref{fraction} that at low ionization fraction $x(r)$ $\sim$ $\sqrt{U}$. As He II-ionizing photons are lost to ionizing Hydrogen, the luminosity of any He II line should fall accordingly (therefore at sufficiently low U, $L_{4686}$ $\sim$ $\rm{U}^{0.5}$). Inspecting the results for varying U in Fig. \ref{He II_dis}, we see that this holds for log(U) $\lesssim$ -4, above which it levels off and saturates. For high source temperatures, we see that the line luminosity falls short of that predicted by eq. \ref{CaseB}, even for log(U) $\approx$ 0. This deficiency grows roughly linearly with increasing $\dot Q_{>54.4\rm{eV}}/\dot Q_{>13.6\rm{eV}}$. This is precisely the effect found by \citet{ST86}, in its original context. As the ratio $\dot Q_{>54.4\rm{eV}}/\dot Q_{>13.6\rm{eV}}$ grows with source temperature, an increasing fraction of photons are lost in maintaining H-ionization. Note that at the same time, the equilibrium temperature for the gas rises, slightly decreasing $\alpha^{\rm{eff}}_{\rm{He II }4\rightarrow 3}(\rm{T_{\rm{gas}}})/\alpha _{\rm{B}}(\rm{He II, T_{\rm{gas}}})$, while increasing $\alpha _{\rm{B}}(\rm{He II, T_{\rm{gas}}})/\alpha _{\rm{B}}(\rm{H I, T_{\rm{gas}}})$ ($\sim$ 6 at $10^{4}$K).

The competition between H- and He-ionization makes it rather difficult to constrain the emission from very high temperature sources, or where the ionization parameter is quite low. At such temperatures there should be a significant enhancement in lines characteristic of ionization by very hard spectra, such as [O I] 6300\AA\ and [N I] 5200\AA\ \citep[see e.g.][]{Rappaport94}, both of which trace the width of the Str{\"o}mgren boundary. The ability to make a robust statement based on such diagnostics would however be far more susceptible to the uncertainties in our knowledge of the warm gas in ellipticals, such as the metallicity and ionization state. As discussed above, measurements of the soft X-ray emission have already placed strong limits in this temperature range, therefore we focus our attention on sources with T $\lesssim$ 6 $\cdot$ $10^{5}K$; this includes the USSs and most SSSs.

\bsp

\label{lastpage}


\begin{thebibliography}{56}
\expandafter\ifx\csname natexlab\endcsname\relax\def\natexlab#1{#1}\fi

\bibitem[{{Anders} \& {Grevesse}(1989)}]{AG89}
{Anders} E., {Grevesse} N., 1989, Geochimica et Cosmochimica Acta, 53, 197

\bibitem[{{Annibali} {et~al}\mbox{.}(2007){Annibali}, {Bressan}, {Rampazzo},
  {Zeilinger}, \& {Danese}}]{Annibali07}
{Annibali} F., {Bressan} A., {Rampazzo} R., {Zeilinger} W.~W., {Danese} L.,
  2007, Astronomy and Astrophysics, 463, 455

\bibitem[{{Annibali} {et~al}\mbox{.}(2010){Annibali}, {Bressan}, {Rampazzo},
  {Zeilinger}, {Vega}, \& {Panuzzo}}]{Annibali10}
{Annibali} F., {Bressan} A., {Rampazzo} R., {Zeilinger} W.~W., {Vega} O.,
  {Panuzzo} P., 2010, Astronomy and Astrophysics, 519, A40

\bibitem[{{Athey} \& {Bregman}(2009)}]{AB09}
{Athey} A.~E., {Bregman} J.~N., 2009, The Astrophysical Journal, 696, 681

\bibitem[{{Binette} {et~al}\mbox{.}(1994){Binette}, {Magris}, {Stasi{\'n}ska},
  \& {Bruzual}}]{Binette94}
{Binette} L., {Magris} C.~G., {Stasi{\'n}ska} G., {Bruzual} A.~G., 1994,
  Astronomy and Astrophysics, 292, 13

\bibitem[{{Brown} {et~al}\mbox{.}(2008){Brown}, {Smith}, {Ferguson},
  {Sweigart}, {Kimble}, \& {Bowers}}]{Brown08}
{Brown} T.~M., {Smith} E., {Ferguson} H.~C., {Sweigart} A.~V., {Kimble} R.~A.,
  {Bowers} C.~W., 2008, The Astrophysical Journal, 682, 319

\bibitem[{{Bruzual} \& {Charlot}(2003)}]{BC03}
{Bruzual} G., {Charlot} S., 2003, Monthly Notices of The Royal Astronomical
  Society, 344, 1000

\bibitem[{{Buyle} {et~al}\mbox{.}(2006){Buyle}, {Michielsen}, {De Rijcke},
  {Pisano}, {Dejonghe}, \& {Freeman}}]{Buyle06}
{Buyle} P., {Michielsen} D., {De Rijcke} S., {Pisano} D.~J., {Dejonghe} H.,
  {Freeman} K., 2006, The Astrophysical Journal, 649, 163

\bibitem[{{Cassisi}, {Iben} \& {Tornambe}(1998){Cassisi}, {Iben}, \&
  {Tornambe}}]{Cassisi98}
{Cassisi} S., {Iben}, Jr. I., {Tornambe} A., 1998, The Astrophysical Journal,
  496, 376

\bibitem[{{Chabrier}(2003)}]{Chabrier}
{Chabrier} G., 2003, Publications of the Astronomical Society of the Pacific,
  115, 763

\bibitem[{{Cid Fernandes} {et~al}\mbox{.}(2011){Cid Fernandes},
  {Stasi{\'n}ska}, {Mateus}, \& {Vale Asari}}]{Fernandes11}
{Cid Fernandes} R., {Stasi{\'n}ska} G., {Mateus} A., {Vale Asari} N., 2011,
  Monthly Notices of the Royal Astronomical Society, 413, 1687

\bibitem[{{Cumming} {et~al}\mbox{.}(1996){Cumming}, {Lundqvist}, {Smith},
  {Pettini}, \& {King}}]{Cumming96}
{Cumming} R.~J., {Lundqvist} P., {Smith} L.~J., {Pettini} M., {King} D.~L.,
  1996, Monthly Notices of the Royal Astronomical Society, 283, 1355

\bibitem[{{Di Stefano}(2010)}]{DiStefano10}
{Di Stefano} R., 2010, Astrophysical Journal, 712, 728

\bibitem[{{Dressler} \& {Gunn}(1983)}]{DG83}
{Dressler} A., {Gunn} J.~E., 1983, The Astrophysical Journal, 270, 7

\bibitem[{{Eracleous}, {Hwang} \& {Flohic}(2010){Eracleous}, {Hwang}, \&
  {Flohic}}]{Eracleous10}
{Eracleous} M., {Hwang} J.~A., {Flohic} H.~M.~L.~G., 2010, Astrophysical
  Journal, 711, 796

\bibitem[{{Gil de Paz} {et~al}\mbox{.}(2007){Gil de Paz}, {Boissier}, {Madore},
  {Seibert}, {Joe}, {Boselli}, {Wyder}, {Thilker}, {Bianchi}, {Rey}, {Rich},
  {Barlow}, {Conrow}, {Forster}, {Friedman}, {Martin}, {Morrissey}, {Neff},
  {Schiminovich}, {Small}, {Donas}, {Heckman}, {Lee}, {Milliard}, {Szalay}, \&
  {Yi}}]{GALEX_Catalog}
{Gil de Paz} A. {et~al.}, 2007, The Astrophysical Journal Supplement Series,
  173, 185

\bibitem[{{Gilfanov} \& {Bogd{\'a}n}(2010)}]{GB10}
{Gilfanov} M., {Bogd{\'a}n} {\'A}., 2010, Nature, 463, 924

\bibitem[{{Gilfanov} \& {Bogd{\'a}n}(2011)}]{Gilfanov11}
{Gilfanov} M., {Bogd{\'a}n} {\'A}., 2011, ASTROPHYSICS OF NEUTRON STARS 2010: A
  Conference in Honor of M.~Ali Alpar.~ AIP Conference Proceedings, Volume
  1379, pp.~17-22 (2011)., 1379, 17

\bibitem[{{Groves}, {Dopita} \& {Sutherland}(2004){Groves}, {Dopita}, \&
  {Sutherland}}]{Groves04}
{Groves} B.~A., {Dopita} M.~A., {Sutherland} R.~S., 2004, The Astrophysical
  Journal Supplement Series, 153, 9

\bibitem[{{Hachisu}, {Kato} \& {Nomoto}(1999){Hachisu}, {Kato}, \&
  {Nomoto}}]{HKN99}
{Hachisu} I., {Kato} M., {Nomoto} K., 1999, The Astrophysical Journal, 522, 487

\bibitem[{{Hachisu}, {Kato} \& {Nomoto}(2010){Hachisu}, {Kato}, \&
  {Nomoto}}]{HKN10}
{Hachisu} I., {Kato} M., {Nomoto} K., 2010, The Astrophysical Journal Letters,
  724, L212

\bibitem[{{Ho}(2008)}]{Ho08}
{Ho} L.~C., 2008, Annual Reviews of Astronomy and Astrophysics, 46, 475

\bibitem[{{Iben} \& {Tutukov}(1984)}]{IT84}
{Iben}, Jr. I., {Tutukov} A.~V., 1984, The Astrophysical Journals, 54, 335

\bibitem[{{Jarrett} {et~al}\mbox{.}(2003){Jarrett}, {Chester}, {Cutri},
  {Schneider}, \& {Huchra}}]{2MASS}
{Jarrett} T.~H., {Chester} T., {Cutri} R., {Schneider} S.~E., {Huchra} J.~P.,
  2003, The Astronomical Journal, 125, 525

\bibitem[{{Kahabka} \& {van den Heuvel}(1997)}]{Kahabka97}
{Kahabka} P., {van den Heuvel} E.~P.~J., 1997, Annual Review of Astronomy and
  Astrophysics, 35, 69

\bibitem[{{Kaviraj} {et~al}\mbox{.}(2007){Kaviraj}, {Kirkby}, {Silk}, \&
  {Sarzi}}]{Kaviraj07}
{Kaviraj} S., {Kirkby} L.~A., {Silk} J., {Sarzi} M., 2007, Monthly Notices of
  the Royal Astronomical Society, 382, 960

\bibitem[{{Kewley} {et~al}\mbox{.}(2001){Kewley}, {Dopita}, {Sutherland},
  {Heisler}, \& {Trevena}}]{Kewley01}
{Kewley} L.~J., {Dopita} M.~A., {Sutherland} R.~S., {Heisler} C.~A., {Trevena}
  J., 2001, The Astrophysical Journal, 556, 121

\bibitem[{{Lepo} \& {van Kerkwijk}(2011)}]{USS}
{Lepo} K., {van Kerkwijk} M., 2011, ArXiv e-prints

\bibitem[{{Li} {et~al}\mbox{.}(2011){Li}, {Bloom}, {Podsiadlowski}, {Miller},
  {Cenko}, {Jha}, {Sullivan}, {Howell}, {Nugent}, {Butler}, {Ofek}, {Kasliwal},
  {Richards}, {Stockton}, {Shih}, {Bildsten}, {Shara}, {Bibby}, {Filippenko},
  {Ganeshalingam}, {Silverman}, {Kulkarni}, {Law}, {Poznanski}, {Quimby},
  {McCully}, {Patel}, {Maguire}, \& {Shen}}]{Li11}
{Li} W. {et~al.}, 2011, Nature, 480, 348

\bibitem[{{Maggi}(2011)}]{Maggi11}
{Maggi} P., 2011, Master's thesis, Universit\'{e} de Strasbourg, Observatoire
  Astronomique, 11, rue de l'Universit\'{e}, 67000 Strasbourg

\bibitem[{{Mannucci} {et~al}\mbox{.}(2005){Mannucci}, {Della Valle}, {Panagia},
  {Cappellaro}, {Cresci}, {Maiolino}, {Petrosian}, \& {Turatto}}]{Mannucci05}
{Mannucci} F., {Della Valle} M., {Panagia} N., {Cappellaro} E., {Cresci} G.,
  {Maiolino} R., {Petrosian} A., {Turatto} M., 2005, Astronomy and
  Astrophysics, 433, 807

\bibitem[{{Maoz} \& {Mannucci}(2012)}]{MM12}
{Maoz} D., {Mannucci} F., 2012, Publications of the Astronomical Society of
  Australia, 29, 447

\bibitem[{{Matteucci} \& {Greggio}(1986)}]{MG86}
{Matteucci} F., {Greggio} L., 1986, Astronomy and Astrophysics, 154, 279

\bibitem[{{Nielsen}, {Voss} \& {Nelemans}(2012){Nielsen}, {Voss}, \&
  {Nelemans}}]{NVN12}
{Nielsen} M.~T.~B., {Voss} R., {Nelemans} G., 2012, Monthly Notices of the
  Royal Astronomical Society, 426, 2668

\bibitem[{{Nomoto} {et~al}\mbox{.}(2007){Nomoto}, {Saio}, {Kato}, \&
  {Hachisu}}]{Nomoto07}
{Nomoto} K., {Saio} H., {Kato} M., {Hachisu} I., 2007, The Astrophysical
  Journal, 663, 1269

\bibitem[{{Osterbrock}(1989)}]{Osterbrock}
{Osterbrock} D.~E., 1989, {Astrophysics of gaseous nebulae and active galactic
  nuclei}

\bibitem[{{Perlmutter} {et~al}\mbox{.}(1999){Perlmutter}, {Aldering},
  {Goldhaber}, {Knop}, {Nugent}, {Castro}, {Deustua}, {Fabbro}, {Goobar},
  {Groom}, {Hook}, {Kim}, {Kim}, {Lee}, {Nunes}, {Pain}, {Pennypacker},
  {Quimby}, {Lidman}, {Ellis}, {Irwin}, {McMahon}, {Ruiz-Lapuente}, {Walton},
  {Schaefer}, {Boyle}, {Filippenko}, {Matheson}, {Fruchter}, {Panagia},
  {Newberg}, {Couch}, \& {Supernova Cosmology Project}}]{Perlmutter99}
{Perlmutter} S. {et~al.}, 1999, The Astrophysical Journal, 517, 565

\bibitem[{{Raiter}, {Schaerer} \& {Fosbury}(2010){Raiter}, {Schaerer}, \&
  {Fosbury}}]{RSF10}
{Raiter} A., {Schaerer} D., {Fosbury} R.~A.~E., 2010, Astronomy and
  Astrophysics, 523, A64

\bibitem[{{Rappaport} {et~al}\mbox{.}(1994){Rappaport}, {Chiang}, {Kallman}, \&
  {Malina}}]{Rappaport94}
{Rappaport} S., {Chiang} E., {Kallman} T., {Malina} R., 1994, The Astrophysical
  Journal, 431, 237

\bibitem[{{Rauch}(2003)}]{Rauch03}
{Rauch} T., 2003, Astronomy and Astrophysics, 403, 709

\bibitem[{{Rauch} \& {Werner}(2010)}]{Rauch10}
{Rauch} T., {Werner} K., 2010, Astronomische Nachrichten, 331, 146

\bibitem[{{Riess} {et~al}\mbox{.}(1998){Riess}, {Filippenko}, {Challis},
  {Clocchiatti}, {Diercks}, {Garnavich}, {Gilliland}, {Hogan}, {Jha},
  {Kirshner}, {Leibundgut}, {Phillips}, {Reiss}, {Schmidt}, {Schommer},
  {Smith}, {Spyromilio}, {Stubbs}, {Suntzeff}, \& {Tonry}}]{Riess98}
{Riess} A.~G. {et~al.}, 1998, The Astronomical Journal, 116, 1009

\bibitem[{{Ruiter}, {Belczynski} \& {Fryer}(2009){Ruiter}, {Belczynski}, \&
  {Fryer}}]{Ruiter09}
{Ruiter} A.~J., {Belczynski} K., {Fryer} C., 2009, Astrophysical Journal, 699,
  2026

\bibitem[{{S{\'a}nchez} {et~al}\mbox{.}(2012){S{\'a}nchez}, {Kennicutt}, {Gil
  de Paz}, {van de Ven}, {V{\'{\i}}lchez}, {Wisotzki}, {Walcher}, {Mast},
  {Aguerri}, {Albiol-P{\'e}rez}, {Alonso-Herrero}, {Alves}, {Bakos},
  {Bart{\'a}kov{\'a}}, {Bland-Hawthorn}, {Boselli}, {Bomans},
  {Castillo-Morales}, {Cortijo-Ferrero}, {de Lorenzo-C{\'a}ceres}, {Del Olmo},
  {Dettmar}, {D{\'{\i}}az}, {Ellis}, {Falc{\'o}n-Barroso}, {Flores},
  {Gallazzi}, {Garc{\'{\i}}a-Lorenzo}, {Gonz{\'a}lez Delgado}, {Gruel},
  {Haines}, {Hao}, {Husemann}, {Igl{\'e}sias-P{\'a}ramo}, {Jahnke}, {Johnson},
  {Jungwiert}, {Kalinova}, {Kehrig}, {Kupko}, {L{\'o}pez-S{\'a}nchez},
  {Lyubenova}, {Marino}, {M{\'a}rmol-Queralt{\'o}}, {M{\'a}rquez}, {Masegosa},
  {Meidt}, {Mendez-Abreu}, {Monreal-Ibero}, {Montijo}, {Mour{\~a}o},
  {Palacios-Navarro}, {Papaderos}, {Pasquali}, {Peletier}, {P{\'e}rez},
  {P{\'e}rez}, {Quirrenbach}, {Rela{\~n}o}, {Rosales-Ortega}, {Roth},
  {Ruiz-Lara}, {S{\'a}nchez-Bl{\'a}zquez}, {Sengupta}, {Singh}, {Stanishev},
  {Trager}, {Vazdekis}, {Viironen}, {Wild}, {Zibetti}, \& {Ziegler}}]{CALIFA}
{S{\'a}nchez} S.~F. {et~al.}, 2012, Astronomy and Astrophysics, 538, A8

\bibitem[{{Sarzi}, {Shields} \& {Schawinski}(2010){Sarzi}, {Shields}, \&
  {Schawinski}}]{SAURON}
{Sarzi} M., {Shields} J.~C., {Schawinski} K.~e.~a., 2010, Monthly Notices of
  the Royal Astronomical Society, 402, 2187

\bibitem[Sarzi et al.(2006)]{SAURON06} Sarzi, M., 
Falc{\'o}n-Barroso, J., Davies, R.~L., et al.\ 2006, Monthly Notices of the Royal Astronomical Society, 366, 1151 

\bibitem[{{Serra} {et~al}\mbox{.}(2012){Serra}, {Oosterloo}, {Morganti},
  {Alatalo}, {Blitz}, \& {Bois}}]{ATLAS3D_12}
{Serra} P., {Oosterloo} T., {Morganti} R., {Alatalo} K., {Blitz} L., {Bois} M.
  e.~a., 2012, Monthly Notices of the Royal Astronomical Society, 422, 1835

\bibitem[{{Stasi{\'n}ska} \& {Tylenda}(1986)}]{ST86}
{Stasi{\'n}ska} G., {Tylenda} R., 1986, Astronomy and Astrophysics, 155, 137

\bibitem[{{Stasi{\'n}ska} {et~al}\mbox{.}(2008){Stasi{\'n}ska}, {Vale Asari},
  {Cid Fernandes}, {Gomes}, {Schlickmann}, {Mateus}, {Schoenell}, {Sodr{\'e}},
  \& {Seagal Collaboration}}]{Stasinska08}
{Stasi{\'n}ska} G. {et~al.}, 2008, Monthly Notices of the Royal Astronomical
  Society, 391, L29

\bibitem[{{Totani} {et~al}\mbox{.}(2008){Totani}, {Morokuma}, {Oda}, {Doi}, \&
  {Yasuda}}]{Totani08}
{Totani} T., {Morokuma} T., {Oda} T., {Doi} M., {Yasuda} N., 2008, Publications
  of the Astronomical Society of Japan, 60, 1327

\bibitem[{{Umeda} {et~al}\mbox{.}(1999){Umeda}, {Nomoto}, {Yamaoka}, \&
  {Wanajo}}]{Umeda99}
{Umeda} H., {Nomoto} K., {Yamaoka} H., {Wanajo} S., 1999, The Astrophysical
  Journal, 513, 861

\bibitem[{{van den Heuvel} {et~al}\mbox{.}(1992){van den Heuvel},
  {Bhattacharya}, {Nomoto}, \& {Rappaport}}]{vdHeuvel92}
{van den Heuvel} E.~P.~J., {Bhattacharya} D., {Nomoto} K., {Rappaport} S.~A.,
  1992, Astronomy and Astrophysics, 262, 97

\bibitem[{{Webbink}(1984)}]{Webbink84}
{Webbink} R.~F., 1984, Astrophysical Journal, 277, 355

\bibitem[{{Whelan} \& {Iben}(1973)}]{WI73}
{Whelan} J., {Iben}, Jr. I., 1973, The Astrophysical Journal, 186, 1007

\bibitem[{{Yagi} \& {Goto}(2006)}]{YG06}
{Yagi} M., {Goto} T., 2006, The Astronomical Journal, 131, 2050

\bibitem[{{Yan} \& {Blanton}(2012)}]{Yan12}
{Yan} R., {Blanton} M.~R., 2012, The Astrophysical Journal, 747, 61

\bibitem[{{Yungelson}(2010)}]{Yungelson10}
{Yungelson} L.~R., 2010, Astronomy Letters, 36, 780

\end{thebibliography}
\end{document}